\input psfig
\documentstyle[aps,preprint]{revtex}

\begin{document}
\draft
\date{\today}
\title{
Singular Pair Breaking in the Composite Fermi Liquid Description of
the Half-Filled Landau Level}
\author{N.\ E. Bonesteel}
\address{
National High Magnetic Field Laboratory and Department of Physics,
Florida State University, Tallahassee, FL 32310}
\maketitle
\begin{abstract}
Fluctuations of the Chern-Simons gauge field in the composite Fermi
liquid description of the half-filled Landau level are pair breaking
in all angular momentum channels.  For short-range electron-electron
interactions these fluctuations are sufficiently strong to drive any
$T=0$ pairing transition first order. For Coulomb interactions these
fluctuations are weaker and a continuous transition is possible.
\end{abstract}

There is strong experimental support \cite{experiments} for the
remarkable hypothesis that the two-dimensional electron gas at
Landau-level filling fraction $\nu=1/2$ can be viewed as a
compressible `metal' with a sharp Fermi surface
\cite{halperinleeread}.  The initial formulation of this idea, based
on the composite fermion theory of the fractional quantum Hall effect
\cite{jain}, involved representing physical electrons by fermions,
referred to in what follows as Chern-Simons (CS) fermions, bound to
two quanta of fictitious, or CS, flux
\cite{halperinleeread,kalmeyer,lopez}.  If this flux is chosen to
point in the direction opposite to that of the physical magnetic field
then, at the mean-field level, CS fermions at $\nu=1/2$ see zero
effective field and form a metallic state with a Fermi surface.
Halperin, Lee, and Read (HLR) \cite{halperinleeread} developed a
theoretical description of the resulting `composite Fermi liquid'
(CFL) by studying fluctuations about this mean-field state within the
random-phase approximation (RPA).

Unlike the CS theories used to describe incompressible fractional
quantum Hall states \cite{lopez,zhang}, where the presence of an
energy gap provides some justification for thinking that fluctuations
are under control, the CFL is gapless, the fluctuations are large, and
the degree to which the mean-field solution captures the essential
physics is unclear.  One question with important physical consequences
is whether or not the mean-field Fermi surface is stable.  Greiter,
Wen, and Wilczek (GWW) \cite{greiter} have shown that the bare
`density-current' interaction between the flux attached to one CS
fermion and the current of another CS fermion mediates an attractive
pairing interaction in the $p$-wave channel.  These authors argued
further that the resulting paired state corresponds to the
incompressible Pfaffian state originally proposed by Moore and Read
\cite{mooreread}.

The fractional quantum Hall state at $\nu = 5/2$ \cite{willett}
provides experimental motivation for studying the stability of the
CFL.  Recent exact diagonalization calculations of Morf \cite{morf}
have shown the for the Coulomb interaction the half-filled
first-excited Landau level at $\nu=5/2$ is spin polarized, even in the
absence of Zeeman coupling.  Morf further speculates that the observed
collapse of this state in tilted fields \cite{eisenstein} is a
consequence of the `hardening' of the short-range part of the
electron-electron interaction due to the `pinching' of the lowest
subband wave function of the two dimensional electron gas \cite{chak}.
If this is the case then it is plausible that the incompressible state
at small tilt is a Pfaffian, and the tilted field transition is from
Pfaffian to CFL.  Another scenario, advocated by Rezayi and Haldane
\cite{hr-aps}, is that for a pure system there is no transition ---
the CFL is {\it always} unstable to the Pfaffian --- and the observed
incompressible-compressible transition occurs when the gap becomes
smaller than the characteristic decay width associated with the
disorder.

The analysis of GWW favors the latter scenario, that in the absence of
disorder the CFL is always unstable to the formation of a paired
quantum Hall state.  It is the modest purpose of this paper to show
that while this may in fact be the case it is not {\it necessarily}
the case.  If one goes beyond GWW and, following
\cite{halperinleeread}, computes the effective interaction between CS
fermions within the RPA, the {\it current-current} interaction
mediated by the transverse CS gauge fluctuations is more singular than
the density-current interaction considered by GWW and is strongly {\it
pair breaking} in all angular momentum channels.  These fluctuations
can, in principle, stabilize the CFL.  In what follows it will be
assumed, with the usual provisos regarding renormalization of the
effective mass, that the HLR approach is qualitatively correct.  The
relevance of the present work to more recent theoretical formulations
of the CFL \cite{read1,baskaran,shankar,pasquier,lee,read2} will
depend on the extent to which these theories resemble HLR, currently a
matter of some controversy \cite{halperinmurthy}.

Consider a two dimensional electron gas, realized in the $xy$ plane,
in a perpendicular magnetic field $B$ at filling factor
$\nu=1/\tilde\phi$, where $\tilde\phi$ is an even integer.  Taking
$\hbar = c = 1$ the magnetic field is $B = 2\pi\tilde\phi n/e$ where
$n$ is the electron density. This system can be described by the
Euclidean time action \cite{halperinleeread} ${\cal S} = \int_0^\beta
d\tau {\cal L}_0(\tau) + {\cal S}_{CS}$, where
\begin{eqnarray}
{\cal L}_0(\tau) =
\int d^2 r~ {\overline \psi}
\left(
\partial_\tau - a_0 - \mu
 - \frac{1}{2 m}
({\bf \nabla} - i {\bf a} + ie {\bf A})^2 \right)
\psi
\end{eqnarray}
and
\begin{eqnarray}
{\cal S}_{CS} = \frac{1}{2}\sum_{\mu,\nu} \sum_{{\bf q},n} a_\mu({\bf
q},\omega_n) {{\cal D}^0_{\mu\nu}}^{-1}(q) a_\nu({\bf q},\omega_n).
\end{eqnarray}
Here $\psi$ is the CS fermion field, $a_0$ and $a_1({\bf q},\omega_n)
= \hat {\bf z}
\cdot (\hat {\bf q} \times {\bf a}({\bf q},\omega_n))$ are the time
and transverse components of the corresponding statistical gauge
field, $\omega_n = 2n\pi/\beta$ is a bosonic Matsubara frequency,
${\bf A}({\bf r}) = ({\hat {\bf z}}\times{\bf r})B/2$ is the physical
vector potential describing the applied magnetic field, $m$ is the
band mass of the electrons, and
\begin{eqnarray}
{\cal D}^0(q) =
\left(
\begin{array}{cc}
 v(q)  & i2\pi \tilde \phi/q \\
-i 2\pi\tilde \phi/q & 
0
\end{array}
\right)
\end{eqnarray}
is the `bare' CS propagator where $v(q)$ is the electron-electron
interaction.  Integrating out $a_0$ enforces the constraint
${\bbox{\nabla}}\times{\bf a} = \hat {\bf z} 2\pi\tilde\phi {\overline
\psi} \psi$ so that at the mean-field level
$\langle{\bbox{\nabla}}\times{\bf a}\rangle = \hat {\bf z}
2\pi\tilde\phi n = \hat {\bf z} e B$.  The CS gauge field then exactly
cancels the physical magnetic field and the CS fermions form a Fermi
liquid with Fermi wave vector $k_F = (2/\tilde\phi)^{1/2}/l_0$ where
$l_0 = (1/eB)^{1/2}$ is the magnetic length.


The pairing instability discussed by GWW \cite{greiter} is due to the
bare `statistical' interaction in the Cooper channel,
\begin{eqnarray}
V^0_{10}({\bf k},{\bf k}^\prime) = \frac{{\bf k} \times {\bf \hat
q}}{m} {\cal D}^0_{10}(q) = i \frac{{\bf k} \times {\bf \hat
q}}{m}\frac{2\pi\tilde\phi}{q},
\end{eqnarray}
where ${\bf q} = {\bf k} - {\bf k}^\prime$.  A dimensionless coupling
constant characterizing the strength of a given pairing interaction
can be obtained by taking the Fermi-surface average in the $l$-wave
channel,
\begin{eqnarray}
\lambda(\omega) = 
-\frac{m}{2\pi}
\int_0^{2\pi} 
V\left(k_F {\hat {\bf x}},k_F(\cos\theta \hat {\bf x} + \sin\theta
\hat {\bf y}) ,\omega\right)
\exp(i \theta l) d \theta,
\label{cc}
\end{eqnarray}
where the sign is chosen so that $\lambda$ is positive for an
attractive interaction.  Because the CS fermions are spinless $l$ must
be odd.  For $V^0_{10}$ there is no frequency dependence and the
result of this integral is $\lambda^0_{10} = {\rm sgn}(l)
\pi\tilde\phi$ \cite{fsapprox}.  The interaction is attractive for
positive $l$ and repulsive for negative $l$, reflecting the fact that
the pairing interaction is not time-reversal symmetric.

The RPA expression for the CS gauge field propagator is obtained by
integrating out the CS Fermi fields and expanding the resulting
effective action to second order in the CS gauge fields with the
result
\cite{halperinleeread}
\begin{eqnarray}
{\cal S}_{RPA} = \frac{1}{2}\sum_{\mu,\nu}\sum_{{\bf q},n} a_\mu({\bf
q},\omega_n) {\cal D}_{\mu\nu}^{-1}(q,\omega_n) a_\nu({\bf
q},\omega_n).
\end{eqnarray}
Here ${\cal D}^{-1} = {\cal K}^0 + {{\cal D}^0}^{-1}$ where ${\cal
K}_{\mu\nu}^0$ is the electromagnetic response function for non
interacting electrons \cite{halperinleeread}.  For $q < 2 k_F$ and
$\omega \ll k_F q/m$, ${\cal K}^0_{00} \simeq m/2\pi$ and ${\cal
K}^0_{11} \simeq -\chi_{\rm d} q^2 - k_F|\omega|/4\pi q$, where
$\chi_{d} = (12\pi m)^{-1}$ is the Landau diamagnetic susceptibility.

\begin{figure}
\centerline{
\psfig{figure=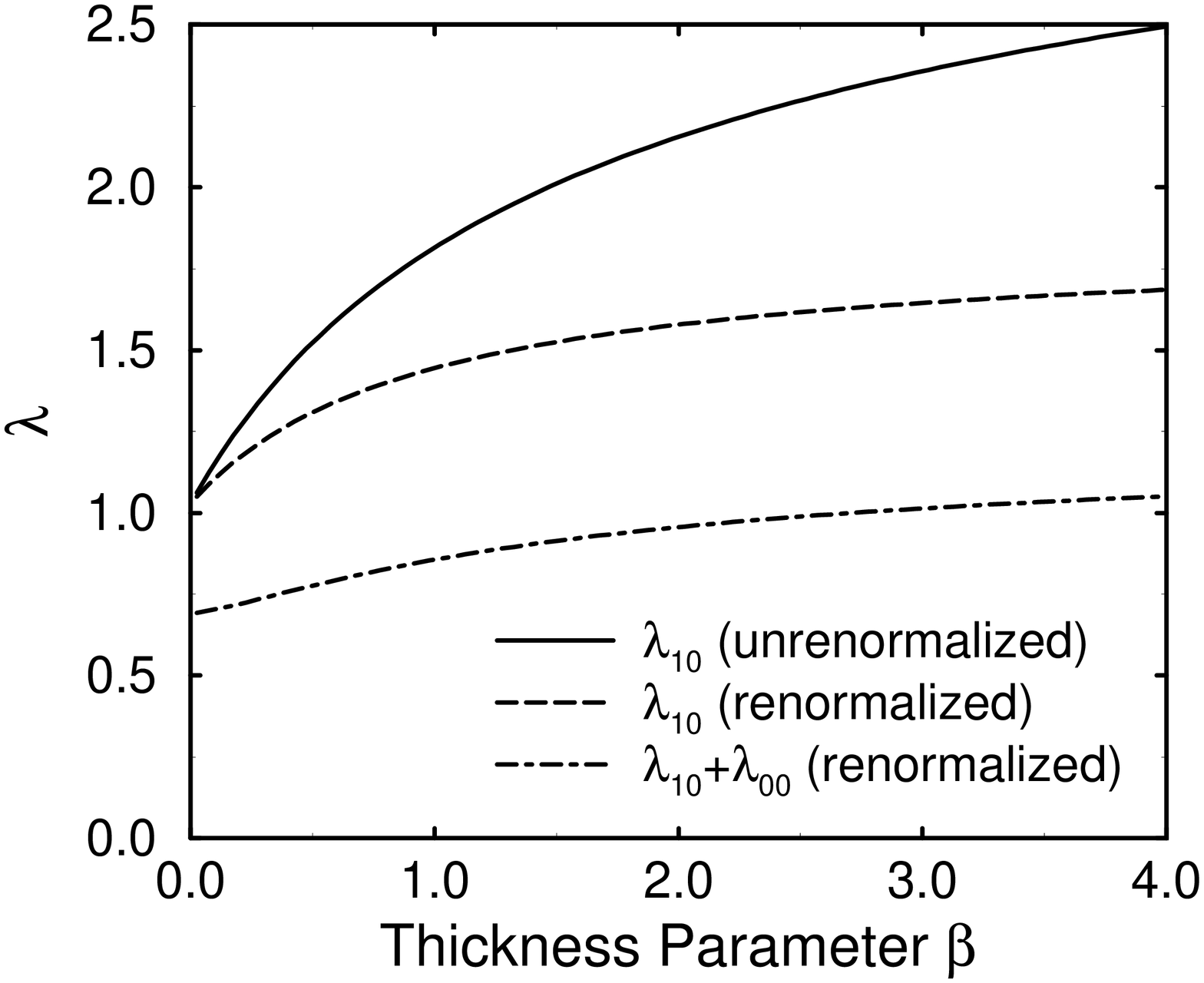,height=5in,angle=-90}}
\
\
\caption{
Coupling constant characterizing the strength of the $p$-wave pairing
interaction vs.\ thickness parameter $\beta$.  Results are given with
(solid) and without (dashed) the mass renormalization discussed in the
text, as well as with both the mass renormalization and the screened
Coulomb repulsion (dot-dashed).  In all cases the interaction grows
stronger as the thickness is increased.}
\label{lambdacs}
\end{figure}

Within the RPA the {\it screened} density-current interaction in the
static limit is
\begin{eqnarray}
V^{RPA}_{10}({\bf k},{\bf k^\prime},\omega = 0) =
\frac{V_{10}^0({\bf k},{\bf k}^\prime)}
{2\pi m \tilde\phi^2 \tilde \chi(q)}
\end{eqnarray}
where $\tilde\chi(q) = v(q)/(2\pi\tilde\phi)^2+(1+6/\tilde
\phi^2)/(12\pi m)$.  Before proceeding it is interesting to study
the dependence of the strength of this pairing interaction on the
`thickness' of the two dimensional electron gas. This can be done
using the Stern-Howard potential for $v(q)$ corresponding to subband
wave function $\xi(w) = A w \exp -bw/2$ \cite{stern}.  In order to
address the well known failure of the various CS theories to
renormalize the bare mass $m$ so that it is determined by $v(q)$, the
dependence of the energy gap of the $\nu=1/3$ fractional quantum Hall
state on the parameter $\beta = (b l_0)^{-1}$, computed variationally
in \cite{melik}, has been used.  The results of these calculations are
well fit by the simple expression $\Delta_{1/3}(\beta) \simeq
0.1/(1+0.7\beta) e^2/l_0$ which, using the relation
$\Delta_{1/3}(\beta) = eB_{eff}/mc=1/3ml_0^2$, gives the $\beta$
dependent effective mass, $m(\beta) \simeq (10+7\beta)/(3e^2 l_0)$.
Figure \ref{lambdacs} shows the dependence of $\lambda_{10}(0)$ on the
thickness parameter $\beta$, computed using (\ref{cc}) with $l=1$ and
$\tilde \phi=2$, both with and without this mass renormalization.  The
effect of including the screened density-density interaction
$V_{00}({\bf q},\omega=0) = (v_q+(2\pi\tilde\phi)^2/(12\pi m))/(2\pi m
\tilde\phi^2\tilde
\chi(q))$ through the corresponding coupling constant
$\lambda_{00}(0)$ is also shown.  In all cases the effective $p$-wave
pairing interaction grows with thickness.

So far the current-current interaction between CS fermions mediated by
the transverse CS gauge fluctuations has been ignored. In the Cooper
channel this interaction is
\begin{eqnarray}
V_{11}^{RPA}({\bf k},{\bf k}^\prime,\omega_n) =
\left(\frac{{\bf k} \times {\bf \hat q}}{m}\right)^2 
{\cal D}_{11}(q,\omega_n)
\simeq 
\left(\frac{{\bf k} \times {\bf \hat q}}{m}\right)^2
\frac{1}{\tilde \chi(q) q^2 + (k_F/2\pi)|\omega_n|/q}.
\end{eqnarray}
It is convenient to first consider the case of short-range
electron-electron interactions.  Taking $v(q) \simeq v(0)$ and
evaluating (\ref{cc}) gives
\begin{eqnarray}
\lambda_{11}(\omega_n) \sim -
\frac{1}{m}\left(\frac{k_F}{\tilde\chi(0)}\right)^{2/3} 
|\omega_n|^{-1/3}
\end{eqnarray}
Because the divergence of $\lambda_{11}(\omega)$ as $\omega
\rightarrow 0$ comes from small $q$ scattering of Cooper pairs it is
independent of $l$.  Note that this singularity is {\it repulsive} and
hence pair breaking in all angular momentum channels.

The effect of this singularity on the pairing instability of the CFL
can be seen by considering the following simplified $T=0$ BCS gap
equation,
\begin{eqnarray}
\Delta(\omega)\, &=&\,\lambda \int^{\omega_0}_{-\omega_0} d\omega^\prime 
\frac{\Delta(\omega^\prime)}{2 \sqrt{{\omega^\prime}^2 + |\Delta(\omega^\prime)|^2}}
- \gamma \int_{-\infty}^{\infty}d\omega^\prime
\frac{\Delta(\omega^\prime)}{2\sqrt{{\omega^\prime}^2
+|\Delta(\omega^\prime)|^2}}
\left(\frac{\omega_0}{|\omega - \omega^\prime|}\right)^{1/3}.
\end{eqnarray}
Here $\lambda$ and $\gamma$ are dimensionless coupling constants
characterizing a nonsingular attractive interaction and a singular
repulsive interaction and $\omega_0$ is a characteristic energy scale
(presumably of order $e^2/l_0$).  Assuming that the $\omega$
dependence of $\Delta$ is weak, taking $\omega = 0$, and performing
the integrals yields
\begin{eqnarray}
1 \simeq \lambda \log \frac{\omega_0}{|\Delta|} - C\gamma
\left(\frac{\omega_0}{|\Delta|}\right)^{1/3}
\label{gapeq}
\end{eqnarray}
where $C \simeq 4.2$.  For small $\Delta$, the second term which
prevents pairing dominates.  This term is present in all angular
momentum channels suggesting that the CFL may be immune from the GWW
instability, or, for that matter, any Kohn-Luttinger-type instability
\cite{kohnluttinger}, at least for short-range electron-electron
interactions.

This analysis leaves out both self-energy effects and the
self-consistent modification of the CS gauge field propagator, both of
which may be important, particularly for the more physically relevant
Coulomb interaction case for which the CS gauge fluctuations lead only
to logarithmic singularities.  An alternative approach which in
principle includes these effects was introduced by Ubbens and Lee
\cite{ubbenslee} in the context of the $U(1)$ gauge-theory description
of the $t$-$J$ model.  In this approach the free energy or, at $T=0$,
the condensation energy, is computed directly within the RPA as a
function of the gap parameter.

To apply the Ubbens-Lee analysis to the present problem it is
necessary to `probe' the CFL by introducing an $l$-wave pairing
interaction by hand.  This interaction is taken to be of the usual
separable form,
\begin{eqnarray}
{\cal S}_{BCS} = \frac{V_0}{{\cal A} \beta}
\sum_{m,n,n^\prime}\sum_{{\bf k},{\bf k}^\prime}
{\overline\gamma}_{\bf k}
\gamma_{\bf k^\prime}^{\phantom{*}}
{\overline \psi}({\bf k},\Omega_n+\omega_m) {\overline \psi}({-{\bf
k}},-\Omega_n) \psi({\bf
k}^\prime,\Omega_{n^\prime}+\omega_m)\psi(-{\bf
k}^\prime,-\Omega_{n^\prime}),
\end{eqnarray}
where $\cal A$ is the area of the system, $\Omega_n = (2n+1)\pi/\beta$
is a fermionic Matsubara frequency, $\gamma_k =
\Theta(\omega_0+\epsilon_k)\Theta(\omega_0-\epsilon_k) \exp
(i\theta_k)$, $\epsilon_k = (k^2-k_F^2)/2m$ and $\theta_k = \arctan
k_y/k_x$.  The Hubbard-Stratonovich decomposition of the BCS
interaction is accomplished by adding the term ${\cal S}_{HS} = \sum_m
{\overline c}(\omega_m) c(\omega_m)$ to the action where $c(\omega_m)
= \sqrt{{\cal A} \beta/V_0}\Delta(\omega_n) + \sqrt{V_0/\cal A \beta}
\sum_n\sum_{\bf k} \gamma_{\bf k} \psi({\bf k},\Omega_n+\omega_m)
\psi(-{\bf k},-\Omega_n)$ \cite{popov}.  The CS fermion fields can
then be integrated out and, within the static approximation,
$\Delta(\omega_n) = \Delta_0 \delta_{n,0}$, the resulting effective
action can be expanded to second order in the CS gauge fields.
Integrating out these fields and taking the $T=0$ limit then yields
the RPA expression for the condensation energy per unit area,
\begin{eqnarray}
E(\Delta_0) = \frac{|\Delta_0|^2}{V_0} - \frac{m}{2\pi}
\int_{-\omega_0}^{\omega_0} 
\left(\sqrt{\epsilon^2 + |\Delta_0|^2} - 
| \epsilon | \right)~ d\epsilon + E_{CS}(\Delta_0) - E_{CS}(0),
\label{conden}
\end{eqnarray}
where
\begin{eqnarray}
E_{CS}(\Delta_0) = \frac{1}{2}\int_{-\infty}^\infty \frac{d\omega}{2\pi} \int
\frac{d^2 q}{(2\pi)^2} \ln \det {\cal D}^{-1}(q,\omega;\Delta_0).
\label{logint}
\end{eqnarray}
Here ${\cal D}_{\mu\nu}(\Delta_0)$ is the CS gauge field propagator
obtained from the equation ${\cal D}^{-1}(\Delta_0) = {\cal
K}(\Delta_0) + {{\cal D}^0}^{-1}$ where now ${\cal
K}_{\mu\nu}(\Delta_0)$ is the electromagnetic response function of the
paired state calculated for fermions described by the Hamiltonian $H =
\sum_{\bf k}( \epsilon_k \psi^\dagger_{\bf k}
\psi_{\bf k} + [\Delta_0 \sum_{\bf k} \gamma_{\bf k}
\psi^\dagger_{\bf k} \psi^\dagger_{-{\bf k}} + {\rm H.c.}])$.
To analyze (\ref{logint}) it is useful to analytically continue to the
real frequency axis, setting $D_{\mu\nu}(q,\nu = i|\omega_n|) = {\cal
D}_{\mu\nu}(q,\omega)$ and $K_{\mu\nu}(q,\nu=|\omega|) = {\cal
K}_{\mu\nu}(q,\omega)$.  The usual deformation of the imaginary
frequency integration in (\ref{logint}) around the branch cut of the
logarithm on the real axis then gives
\cite{halperinleeread,ubbenslee}
\begin{eqnarray}
E_{CS}(\Delta_0) = \int_0^\infty \frac{d\nu}{2\pi}
\int \frac{d^2 q}{(2\pi)^2}
\tan^{-1} 
\frac{{\rm Im} \det D^{-1}(q,\nu;\Delta_0)}
{{\rm Re} \det D^{-1}(q,\nu;\Delta_0)}.
\end{eqnarray}

In the paired state at $T=0$ there is no damping for frequencies $\nu
\le 2|\Delta_0|$ which implies that ${\rm Im}K_{00}(\Delta_0)
= {\rm Im} K_{11}(\Delta_0) = {\rm Re}K_{01}(\Delta_0) = {\rm
Re}K_{10}(\Delta_0) = 0$ and thus ${\rm Im} \det D^{-1}(\Delta_0) = 0$
for $\nu \le 2 |\Delta_0|$.  Following \cite{ubbenslee}
$E_{CS}(\Delta_0) - E_{CS}(0)$ can then be estimated by calculating
the contributions of these frequencies to $-E_{CS}(0)$ which are lost
in the paired state \cite{seekabsolution}.  For the case of
short-range electron-electron interactions, taking $v(q) \simeq v(0)$,
this approximation gives
\begin{eqnarray}
E_{CS}(\Delta_0) - E_{CS}(0) \simeq \int_0^{2|\Delta_0|}
\frac{d\nu}{2\pi}
\int \frac{d^2q}{(2\pi)^2} \tan^{-1} \frac{k_F}{2\pi\tilde\chi(0)}
\frac{\nu}{q^3}
\sim \left(\frac{k_F}{\tilde\chi(0)}\right)^{2/3} |\Delta_0|^{5/3}.
\end{eqnarray}
For small $\Delta_0$ this term, which is consistent with the singular
term appearing in (\ref{gapeq}), will always dominate the condensation
energy and prevent any continuous zero temperature pairing transition.
Thus for short-range electron-electron interactions any pairing
transition of the CFL will necessarily involve a discontinuous jump in
$\Delta_0$ and so be first order.  As pointed out in \cite{ubbenslee},
the appearance in $E(\Delta_0)$ of a nonanalytic term in
$|\Delta_0|^2$ due to finite frequency gauge fluctuations is a quantum
version of the fluctuation driven first-order transition discussed by
Halperin, Lubensky and Ma \cite{halperinlubenskyma}.  A similar
effect, in which a continuous quantum Hall/insulator transition is
driven first-order by a fluctuating CS gauge field, has been studied
by Pryadko and Zhang
\cite{pryadko}.

If the same approximation is applied to the Coulomb interaction case
it is necessary to cut off the momentum integration at $2k_F$ and the
result is
\begin{eqnarray}
E_{CS}(\Delta_0)-E_{CS}(0) \simeq
\int_0^{2|\Delta_0|}\frac{d\nu}{2\pi}
\int_0^{2k_F} \frac{q dq}{2\pi} \tan^{-1} 
\frac{k_F \tilde\phi^2}{e^2} \frac{\nu}{q^2}
\simeq 
\frac{m_c}{2\pi} |\Delta_0|^2 
\ln \frac{2 k_F e^2}{\tilde\phi^2 |\Delta_0|}
\end{eqnarray}
where $m_c = k_F\tilde\phi^2/2\pi e^2$.  For small $\Delta_0$ the
condensation energy is then $E(\Delta_0) \sim |\Delta_0|^2/V_0 -
((m-m_c)/2\pi) |\Delta_0|^2 \ln (\omega_0/|\Delta_0|)$.  If $m - m_c <
0$ pair-breaking dominates and any $T=0$ pairing transition will be
first order.  Alternatively, if $m-m_c >0$ a continuous transition is
possible and the CFL is unstable for arbitrarily weak $V_0$.  It is
interesting to note that according to the mass renormalization scheme
discussed above $m(\beta) - m_c >0$ for $\tilde\phi=2$ suggesting that
for $\nu=1/2$ the CFL is unstable for Coulomb interactions, though it
is unclear to what extent this result can be trusted.

Because the CS contribution to the condensation energy is due to
low-energy, long-wavelength fluctuations, it is unaffected by any
softening of the short-range part of the electron-electron
interaction.  It follows that as the thickness parameter $\beta$
increases both the effective mass $m$ and $\lambda_{10}$, and hence
the tendency towards pairing, increase while the pair-breaking effects
are essentially unchanged.  Thus the CFL becomes less stable with
increasing thickness, consistent with Morf's interpretation of the
$\nu=5/2$ tilted field experiments \cite{morf} as well as numerical
studies of the half-filled Landau level
\cite{greiter,morf,hr-aps,park}.

To summarize, the CS gauge fluctuations in the CFL description of the
half-filled Landau level are strongly pair breaking in all angular
momentum channels and so provide a hostile environment for the
formation of Cooper pairs.  For short-range electron-electron
interactions these fluctuations are sufficiently strong to drive any
$T=0$ pairing transition of the CFL first-order.  For Coulomb
interactions these fluctuations are weaker and, depending on details,
a continuous transition may be possible. To the extent that more
recent formulations of the CFL contain both a pairing interaction
\cite{baskaran} and a $U(1)$ gauge field
\cite{lee,read2,halperinmurthy} the conclusions of this paper should
still be relevant.

Useful discussions with V. Melik-Alaverdian are acknowledged.  This
work was supported by DOE grant No.\ DE-FG02-97ER45639 and the Alfred
P. Sloan Foundation.

\begin{references}

\bibitem{experiments} 
H.W. Jiang {\em et al.}, Phys. Rev. B {\bf 40}, 12013 (1989);
R.L. Willett {\em et al.}, Phys. Rev. Lett. {\bf 65}, 112 (1990);
R.R. Du {\em et al.}, {\em ibid} {\bf 70}, 2944 (1992); W. Kang {\em
et al.}, {\em ibid.} {\bf 71}, 3850 (1993); V.J. Goldman {\em et al.},
{\em ibid.} {\bf 72}, 2065 (1994); J.H. Smet {\em et al.}, {\em ibid.}
{\bf 77}, 2272 (1996); Phys. Rev. B {\bf 56}, 3606 (1997).

\bibitem{halperinleeread}
B.I. Halperin, P.A. Lee, and N. Read, Phys. Rev. B {\bf 47}, 7312
(1993).

\bibitem{jain} 
J.K. Jain, Phys. Rev. Lett. {\bf 63} (1989) 199; Phys. Rev.  B {\bf
40}, 8079 (1989); {\em ibid.} {\bf 41}, 7653 (1990).

\bibitem{kalmeyer} 
V. Kalmeyer and S.C. Zhang, Phys. Rev. B, {\bf 46}, 9889 (1992).

\bibitem{lopez}
A. Lopez and E. Fradkin, Phys. Rev. B {\bf 44}, 5246 (1991).

\bibitem{zhang}
S.C. Zhang, H. Hanson, and S. Kivelson, Phys. Rev. Lett. {\bf 62}, 82
(1989); {\bf 62}, 980 (E) (1989).

\bibitem{greiter} 
M. Greiter, X.G. Wen, and F. Wilczek, Phys. Rev. Lett. {\bf 66}, 3205
(1991); Nucl. Phys. B {\bf 374}, 567 (1992).

\bibitem{mooreread}
G. Moore and N. Read, Nucl. Phys. B {\bf 360}, 362 (1991).

\bibitem{willett} 
R.L. Willett {\em et al.}, Phys. Rev. Lett. {\bf 59}, 1776 (1987).

\bibitem{morf}
R. Morf, Phys. Rev. Lett. {\bf 80}, 1505 (1998).

\bibitem{eisenstein}
J.P. Eisenstein {\em et al.}, Phys. Rev. Lett. {\bf 61}, 997 (1988).

\bibitem{chak} For a discussion of this effect see 
T. Chakraborty and P. Pietil\"ainen, Phys. Rev. B {\bf 39}, 7971
(1989).

\bibitem{hr-aps} E.H. Rezayi and F.D.M. Haldane, Bull. Am. Phys. Soc. {\bf 43}, 655 (1998).

\bibitem{read1}
N. Read, Semicond. Sci. Technol. {\bf 9}, 1859 (1994).

\bibitem{baskaran}
G. Baskaran, Physica B {\bf 212}, 320 (1995).

\bibitem{shankar} R. Shankar and G. Murthy, Phys. Rev. Lett. {\bf 79}, 4437
(1997); G. Murthy and R. Shankar, cond-mat/9709233; cond-mat/9806380.

\bibitem{pasquier}
V. Pasquier and F.D.M. Haldane, cond-mat/9712169.

\bibitem{lee}
D.-H. Lee, Phys. Rev. Lett. {\bf 80}, 4745 (1998).

\bibitem{read2}
N. Read, cond-mat/9804294.

\bibitem{halperinmurthy}
B.I. Halperin and A. Stern, Phys. Rev. Lett. {\bf 80}, 5457 (1998);
G. Murthy and R. Shankar, {\em ibid.} {\bf 80}, 5458 (1998).

\bibitem{fsapprox}
It is an artifact of the Fermi surface approximation that
$\lambda^0_{10}$ is independent of $l$.  In {\protect\cite{greiter}}
the full BCS gap equation was solved and the leading instability found
in the $l=1$ channel.  

\bibitem{stern}
F. Stern and W.E. Howard, Phys. Rev. {\bf 163}, 816 (1967); T. Ando,
A.B. Fowler, and F. Stern, Rev. Mod. Phys. {\bf 54}, 437 (1982).

\bibitem{melik}
V. Melik-Alaverdian and N.E. Bonesteel, Phys. Rev. B {\bf 52}, R17032
(1995).

\bibitem{kohnluttinger}
W. Kohn and J.M. Luttinger, Phys. Rev. Lett. {\bf 15}, 524 (1965).

\bibitem{ubbenslee} 
M.U. Ubbens and P.A. Lee, Phys. Rev. B {\bf 49}, 6853 (1994).

\bibitem{popov} 
See, for example, V.N. Popov {\it Functional Integrals and Collective
Excitations} (Cambridge University Press, Cambridge, 1987).

\bibitem{seekabsolution}
Though this approximation ignores the modification of $K$ for $\nu >
2|\Delta_0|$, it is assumed here that the pair-breaking effects due to
low-frequency fluctuations are adequately captured.

\bibitem{halperinlubenskyma} 
B.I. Halperin, T.C. Lubensky, and S.K. Ma, Phys. Rev. Lett. {\bf 32},
292 (1974).

\bibitem{pryadko} 
L. Pryadko and S.C. Zhang, Phys. Rev. Lett. {\bf 73}, 3282 (1994).

\bibitem{park}
K. Park, V. Melik-Alaverdian, N.E. Bonesteel, and J. Jain, cond-mat/9806271.

\end{references}
\end{document}